\documentclass{ws-ijmpa}

\begin{document}

\markboth{Stefan E. M\"uller}{KLOE results}


\title{KLOE results at the Frascati $\phi$-factory DA$\Phi$NE}

\author{KLOE collaboration\thanks{
A.~Aloisio,
F.~Ambrosino, 
A.~Antonelli,  
M.~Antonelli,  
C.~Bacci, 
M.~Barva,  
G.~Bencivenni,  
S.~Bertolucci,  
C.~Bini,  
C.~Bloise,  
V.~Bocci, 
F.~Bossi, 
P.~Branchini, 
S.~A.~Bulychjov, 
R.~Caloi, 
P.~Campana,  
G.~Capon,  
T.~Capussela, 
G.~Carboni,  
F.~Ceradini, 
F.~Cervelli,  
F.~Cevenini,  
G.~Chiefari,  
P.~Ciambrone, 
S.~Conetti, 
E.~De~Lucia, 
A.~De~Santis,  
P.~De~Simone,  
G.~De~Zorzi, 
S.~Dell'Agnello, 
A.~Denig, 
A.~Di~Domenico, 
C.~Di~Donato, 
S.~Di~Falco, 
B.~Di~Micco, 
A.~Doria, 
M.~Dreucci, 
O.~Erriquez,  
A.~Farilla,  
G.~Felici, 
A.~Ferrari, 
M.~L.~Ferrer,  
G.~Finocchiaro, 
C.~Forti,         
P.~Franzini, 
C.~Gatti,       
P.~Gauzzi, 
S.~Giovannella, 
E.~Gorini,  
E.~Graziani, 
M.~Incagli, 
W.~Kluge,         
V.~Kulikov, 
F.~Lacava,  
G.~Lanfranchi,  
J.~Lee-Franzini,                    
D.~Leone, 
F.~Lu,
M.~Martemianov,
M.~Martini,
M.~Matsyuk,
W.~Mei,                           
L.~Merola,  
R.~Messi, 
S.~Miscetti,  
M.~Moulson, 
S.~M\"uller, 
F.~Murtas,  
M.~Napolitano, 
F.~Nguyen, 
M.~Palutan,            
E.~Pasqualucci, 
L.~Passalacqua,  
A.~Passeri,   
V.~Patera,
F.~Perfetto, 
E.~Petrolo,         
L.~Pontecorvo, 
M.~Primavera,  
P.~Santangelo, 
E.~Santovetti,  
G.~Saracino, 
R.~D.~Schamberger,  
B.~Sciascia, 
A.~Sciubba,
F.~Scuri,  
I.~Sfiligoi, 	 
A.~Sibidanov,
T.~Spadaro, 
E.~Spiriti,  
M.~Testa, 
L.~Tortora,                      
P.~Valente, 
B.~Valeriani, 
G.~Venanzoni, 
S.~Veneziano,       
A.~Ventura,        
R.~Versaci, 
I.~Villella,
G.~Xu.
} 
                \\ presented by Stefan E. M\"uller
                }         

\address{Institut f\"ur Experimentelle Kernphysik, 
        Universit\"at Karlsruhe, \\ 
        Postfach 3640, D-76021 Karlsruhe, Germany \\
	Email: smueller@iekp.fzk.de}


\maketitle


\begin{abstract}
   The KLOE experiment at the Frascati $\phi$-factory DA$\Phi$NE has
   collected about $0.5$ fb$^{-1}$ of data till the end of the year
   2002. These data allow to perform a wide physics program, ranging
   from the physics of charged and neutral kaons to radiative
   $\phi$-decays. Results are presented for the $K_L$ lifetime and 
   the semileptonic
   processes $K_{S,L} \rightarrow \pi e \nu$. From the light meson
   spectroscopy program, results on the decays $\phi \rightarrow
   f_0(980)\gamma , a_0(980)\gamma$ as well as $\phi \rightarrow \eta
   \gamma , \eta' \gamma$ are presented. 
\end{abstract}

\keywords{Scalar mesons; $\eta$-$\eta'$ mixing; Semileptonic kaon decays; 
$V_{us}$.}

\section{Introduction}
DA$\Phi$NE~\cite{DAPHNE} is an $e^+e^-$ collider working at the $\phi$
resonance peak of $1019.5$ MeV. The $\phi$ meson, produced essentially
at rest, decays mostly in charged and neutral kaon pairs, which makes
DA$\Phi$NE an ideal machine for all kinds of kaon physics. Apart from
this, due to the high amount of produced $\phi$ mesons (ca. 1.5
billion $\phi$ with $\sim 450$ pb$^{-1}$), the radiative decays of the
$\phi$ to scalars and pseudoscalars can be studied with unprecedented
precision. With a branching ratio of BR($\phi \to \eta\gamma)\sim 1.3
\%$, about 20 millions $\eta$ mesons have been collected, which is
probably the world's largest sample for $\eta$ mesons.
The KLOE detector consists of a large volume cylindrical drift
chamber~\cite{kloedc} (3.3 m length, 4 m diameter), which is surrounded
by an electromagnetic calorimeter~\cite{kloeemc}. A superconducting coil 
provides a 0.52 T solenoidal magnetic field. 
\section{Light meson spectroscopy with KLOE}
\subsection{Search for the decays $\phi \to $f$_0 \gamma$ and $\phi \to $a$_0 \gamma$}
\begin{figure}
\centerline{\psfig{file=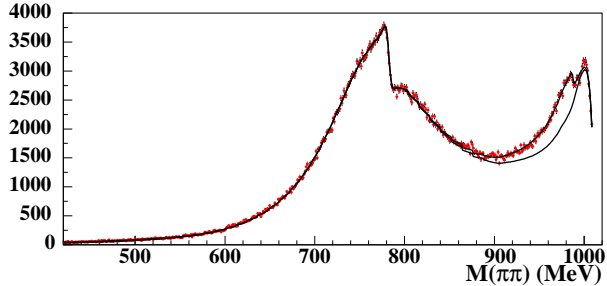,width=8cm}}
\caption{Spectrum for $\pi^+\pi^-\gamma$ events in bins of $M_{\pi\pi}$. A f$_0$ signal peak is visible as a bump around 980 MeV over the background spectrum from initial/final state radiation and $\varrho\pi$ events.}
\label{fig:fig1}
\end{figure}
Apart from the $\pi^0\pi^0\gamma$ final state~\cite{f0neut}, the
f$_0(980)$ signal is also searched for in $\pi^+\pi^-\gamma$ events.
This analysis is made difficult by the presence of a huge irreducible
background from events in which the $e^+$ or the $e^-$ radiate a
photon in the initial state, thus performing a {\em radiative return}
to the $\varrho$ and $\omega$ resonances, which in turn decay to
$\pi^+\pi^-$. To reduce this background, the photon angle for the
signal events is required to be $45^o < \theta_\gamma <
135^o$. Further background arises from $\phi \to \varrho^\pm(\to
\pi^\pm\gamma)\pi^\pm$ events and $e^+e^- \to \pi^+\pi^-\gamma$
events in which the photon is radiated by one of the pions in the
final state. The latter is expected to interfere with the f$_0$
signal.  Fig.~\ref{fig:fig1} shows the obtained spectrum at
$\sqrt{s}=M_\phi$. An overall fit is done to extract the f$_0$
parameters taking into account the different contributions to the
spectrum and also the final state interaction between the two
pions~\cite{ppfi}. The f$_0$ amplitude is modeled using the
kaon-loop approach~\cite{kloop}. 

The decay $\phi \to $a$_0(980)\gamma$ is studied via the $\eta \pi^0
\gamma$ final state, in which the $\eta$ decays either to
$\pi^+\pi^-\pi^0$ or to $\gamma\gamma$. For both decay chains, the
branching ratio for $\phi \to \eta \pi^0 \gamma$ has been obtained
from a fit to the observed spectrum in the $\eta\pi^0$ invariant mass. The 
preliminary results
\begin{eqnarray*}
\eta \to \pi^+\pi^-\pi^0: BR(\phi \to \eta \pi^0 \gamma)&=&(7.45\pm0.19)\cdot10^{-5} \\
\eta \to \gamma\gamma: BR(\phi \to \eta \pi^0 \gamma)&=&(7.25\pm0.15)\cdot10^{-5}
\end{eqnarray*}
compare well with the previous result of 
\mbox{$BR(\phi\to $a$_0\gamma)=(7.4\pm0.7)\cdot10^{-5}$} 
obtained by KLOE~\cite{kloea0}.
\subsection{The decay $\eta \to \gamma\gamma\gamma$ and $\eta$-$\eta'$ mixing}
The decay $\eta \to \gamma\gamma\gamma$ is C-violating, and thus a
sensitive test of C violation in strong and electromagnetic
interactions. The $\eta$ is produced together with a monochromatic
photon of 363 MeV via a radiative decay of the $\phi$. The main
backgrounds to this 4 photon final state are processes which involve
the $\pi^0$ decaying to two photons, so a veto is put on events in
which the invariant mass of any photon pair matches the $\pi^0$
mass. The signal is then looked for in a fit to the distribution of
the energy of the most energetic photon in the event. The upper limit
on the branching ratio for $\eta \to \gamma\gamma\gamma$ is evaluated
by normalizing to the branching ratio for the process $\eta \to
3\pi^0$, taken from the PDG~\cite{pdg}. The KLOE result is~\cite{etaggg}
$BR(\eta \to \gamma\gamma\gamma) < 2.0\cdot10^{-5}@95\%C.L.$

To measure the ratio of the branching ratios for $\phi \to \eta'
\gamma$ and $\phi \to \eta \gamma$, the $\eta'$ is identified via the
decay chains $\phi \to \eta'\gamma$; $\eta' \to \pi^+\pi^-\eta$; $\eta
\to \pi^0\pi^0\pi^0$ and $\phi \to \eta'\gamma$; $\eta' \to
\pi^0\pi^0\eta$; $\eta \to \pi^+\pi^-\pi^0$. The final state is
characterized by two charged pions and seven photons. Normalizing to
the number of observed $\eta \to \pi^0\pi^0\pi^0$ decays in the same
data sample, one obtains a preliminary measurement of the ratio
$R=\frac{BR(\phi \to \eta'\gamma)}{BR(\phi \to \eta\gamma)}=(4.9 \pm
0.1_{stat} \pm 0.2_{syst})\cdot 10^{-3}$. This result compares well
with the previous KLOE estimate~\cite{etaetap}, but with considerably
improved acccuracy.
\section{Neutral kaon decays}
The fact that at KLOE the $K_S$ and the $K_L$ are always produced in
pairs allows for tagged $K_S$, $K_L$ {\em beams} in which one particle
is identified by the presence of the other one. In this way, a $K_S$
can be identified by the characteristic signature of the interaction
of a $K_L$ with the calorimeter, while a $K_L$ is tagged by the decay
of a $K_S \to \pi^+\pi^-$ close to the interaction point.

\begin{figure}
\centerline
{\psfig{file=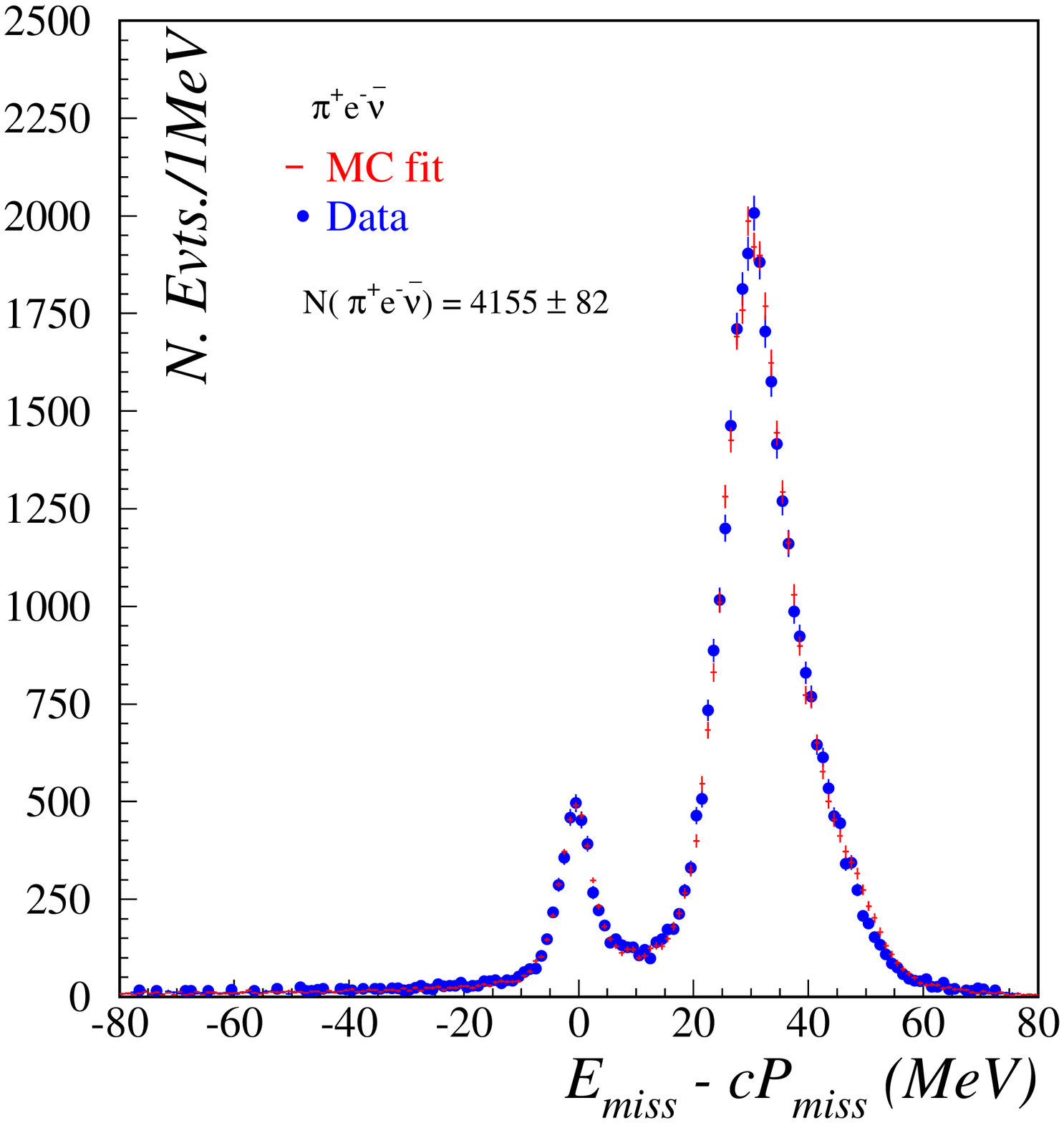,width=5cm,height=4.cm}\psfig{file=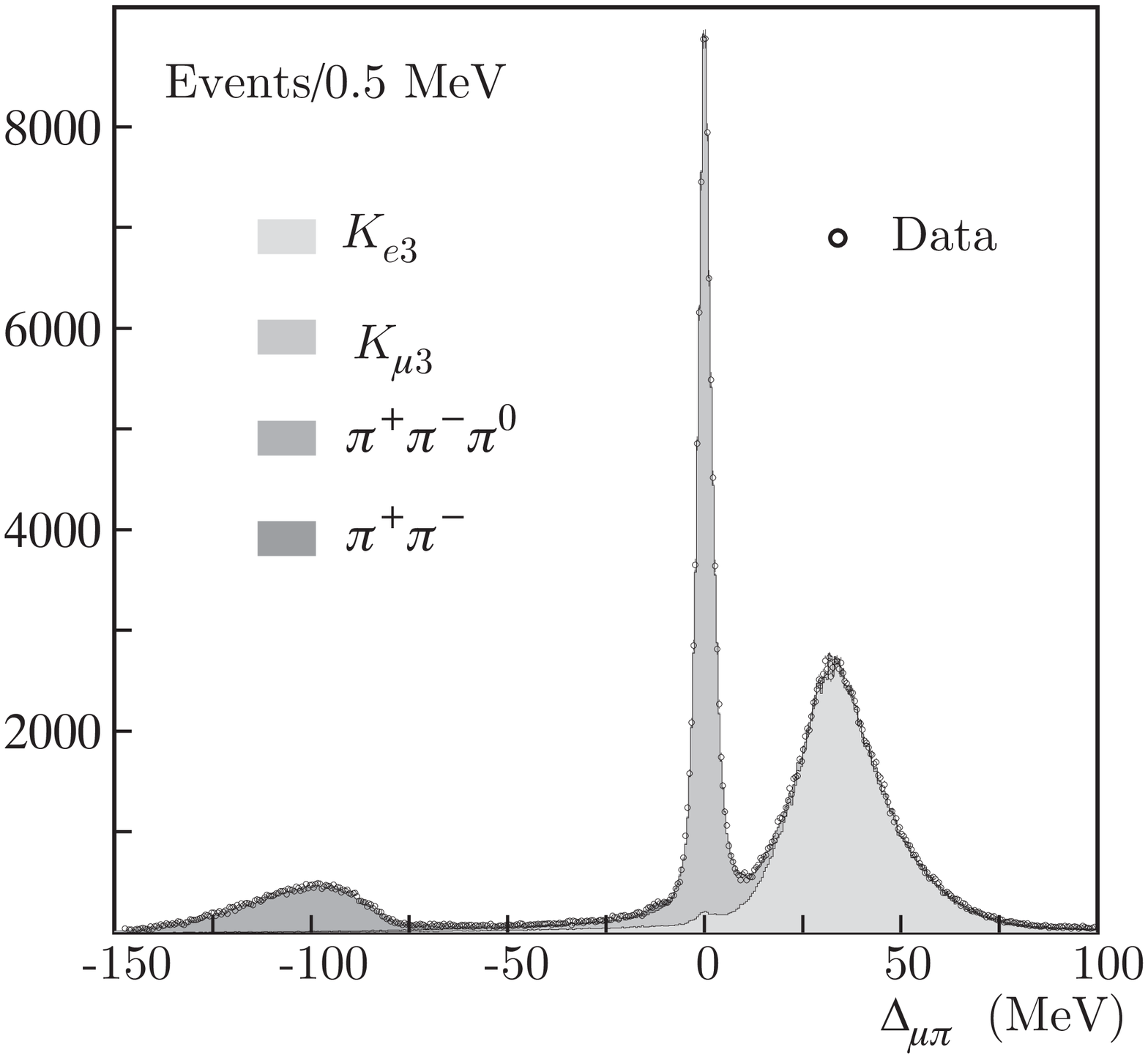,width=5cm,height=3.9cm}}
\caption{Left: $E_{miss}-c\cdot p_{miss}$ for
$\pi^+e^-\overline{\nu}$. Filled dots represent data, crosses
represent the fit result. Right: $\Delta_{\mu\pi}=c\cdot
p_{miss}-E_{miss}$ distribution for data and different MC contributions.}
\label{fig:fig2}
\end{figure}

$K_S \to \pi e \nu$ events are selected by the presence of two
oppositely charged tracks forming a vertex close to the interaction
region. Further cuts are applied to reject most of the background due
to $K_S \to \pi^+\pi^-$ decays. The signal and the remaining
background from $K_S \to \pi^+\pi^-$ (which is used as normalizing
sample) are separated kinematically by building the variable
$E_{miss}-c\cdot p_{miss}$, which peaks at 0 for the signal events
since the missing energy and momentum is carried away by a
neutrino. The number of signal events is then obtained by fitting a
linear combination of Monte Carlo spectra for signal and background to
the data distribution for $E_{miss}-c\cdot p_{miss}$ (see
Fig.~\ref{fig:fig2}, left). Normalizing the number of signal events to the
number of $K_S \to \pi^+\pi^-(\gamma)$ events in the same data set, one
obtains the preliminary results
\begin{eqnarray*}
BR(K_S \to \pi^-e^+\nu)&=&(3.54\pm0.05_{stat}\pm0.05_{syst})\cdot10^{-4} \\
BR(K_S \to \pi^+e^-\overline{\nu})&=&(3.54\pm0.05_{stat}\pm0.04_{syst})\cdot10^{-4} \\
\Rightarrow BR(K_S \to \pi e\nu(\overline{\nu}))&=&(7.09\pm0.07_{stat}\pm0.08_{syst})\cdot10^{-4}
\end{eqnarray*}
From these results one can build the semileptonic charge asymmetry for
the $K_S$, $A_S=\frac{\Gamma(K_S \to \pi^-e^+\nu)-\Gamma(K_S \to
\pi^+e^-\overline{\nu})}{\Gamma(K_S \to \pi^-e^+\nu)+\Gamma(K_S \to
\pi^+e^-\overline{\nu})}=2\Re e(\epsilon)\sim 3\cdot10^{-3}$. One
obtains $A_S=(-2\pm9_{stat}\pm6_{syst})\cdot10^{-3}$, which is the
first measurement of $A_S$ ever done. It is compatible with the
present measurement for $A_L$~\cite{Along}, as expected from CPT
invariance.

13 Mio. $K_L$ decays are tagged by a decay $K_S \to \pi^+\pi^-$. A
decay vertex is searched for along the $K_L$ line of flight using
information from photons (for $K_L \to 3\pi^0$ decay) of from charged
tracks (for $K_L \to \pi e \nu$, $\pi \mu \nu$ and $\pi^+\pi^-\pi^0$
decays). For the charged decays, the channels are separated
kinematically via $c\cdot p_{miss}-E_{miss}$, and the number of signal
events is obtained by fitting the data distribution with a linear
combination of Monte Carlo samples (see Fig.~\ref{fig:fig2}, right). The
preliminary results obtained in this way read
\begin{eqnarray*}
BR(K_L \to \pi e\nu)&=&(0.3994\pm0.0006_{stat}\pm0.0034_{syst}) \\
BR(K_L \to \pi \mu \nu)&=&(0.2708\pm0.0005_{stat}\pm0.0025_{syst}) \\
BR(K_L \to \pi^+\pi^-\pi^0)&=&(0.1271\pm0.0004_{stat}\pm0.0010_{syst}) \\
BR(K_L \to \pi^0\pi^0\pi^0)&=&(0.2014\pm0.0003_{stat}\pm0.0022_{syst})
\end{eqnarray*}
These results have been obtained using the preliminary $K_L$ lifetime
measurement from the decay $K_L \to \pi^0\pi^0\pi^0$:
$\tau(K_L)=(51.15\pm0.20_{stat}\pm0.40_{syst})$ns. Reverting the
argument, one can also try to extract the $K_L$ lifetime from a
unitarity constraint requiring the sum of all $K_L \to X$ decays to be
1. (where the remaining rare $K_L$ branching ratios are taken from the
PDG~\cite{pdg}). The $K_L$ lifetime then becomes $\tau(K_L)=(51.35\pm0.5_{stat}\pm0.26_{syst})$ns.

The most accurate test for CKM matrix unitarity comes from the first
row: $1-\Delta=|V_{ud}|^2+|V_{us}|^2$, where $\Delta$ is expected to
be $\sim 10^{-5}$. $V_{ud}$ is determined precisely from superallowed
nuclear decays and neutron lifetime~\cite{Vud}, while $|V_{us}|$ is
related to the semileptonic decays of kaons via
\begin{displaymath}
\Gamma(K\to \pi l \nu) \sim |V_{us}\cdot f_+^{K\pi}(0)|^2\cdot
S_{ew}\cdot I(\lambda_+)(1+\Delta I(\lambda_+)/2+\delta_{em})^2,
\end{displaymath}
where $f_+^{K\pi}(0)$ is the vector form factor at zero momentum
transfer, $I(\lambda_+)$ is the result of the phase space integration
after factoring out $f_+^{K\pi}(0)$, and $\lambda_+$ describes the
momentum dependence of the form factor. Long distance radiative
corrections are included in the last term in brackets on the right,
while $S_{ew}$ describes the short distance electroweak corrections.
Using the recent KTeV measurement of the $f_+$ dependence on the
momentum transfer~\cite{KTEV} to evaluate the phase space integral,
one gets the preliminary KLOE results
\begin{eqnarray*}
|V_{us}\cdot f_+^{K\pi}(0)|[K_{Se3}] &=& 0.2171\pm0.0017 \\
|V_{us}\cdot f_+^{K\pi}(0)|[K_{Le3}] &=& 0.2147\pm0.0014 \\
|V_{us}\cdot f_+^{K\pi}(0)|[K_{L\mu 3}] &=& 0.2167\pm0.0015
\end{eqnarray*}
which are in good agreement with the value obtained using $f_+^{K\pi}(0)$
from~\cite{LeutRoos} and unitarity (neglecting $V_{ub}$): $(1-|V_{ud}|^2)^{1/2}\cdot f_+^{K\pi}(0)=0.2177\pm0.0028$.


\begin{thebibliography}{0}


\bibitem{DAPHNE} S.~Guiducci, in: Proc. 2001 Particle Accelerator
Conf. (PAC), Chicago (2001) 

\bibitem{kloedc}
KLOE coll., M.~Adinolfi {\it et al.},
Nucl.\ Instrum.\ Meth.\ A {\bf 488} (2002) 51

\bibitem{kloeemc}
KLOE coll., M.~Adinolfi {\it et al.},
Nucl.\ Instrum.\ Meth.\ A {\bf 482} (2002) 364

\bibitem{f0neut}
KLOE coll., A.~Aloisio {\it et al.}, Phys. Lett. B {\bf 537} (2002) 21

\bibitem{ppfi}
N.~N.~Achasov, V.~V.~Gubin, E.~P.~Solodov, Phys. Rev. D {\bf 55} (1997) 2672

\bibitem{kloop}
N.~N.~Achasov, V.~V.~Gubin, Phys. Rev. D {\bf 57} (1998) 1987 

\bibitem{kloea0}
KLOE coll., A.~Aloisio {\it et al.}, Phys. Lett. B {\bf 536} (2002) 209

\bibitem{pdg}
Particle Data Group. Phys. Rev. D {\bf 66} (2002) 1

\bibitem{etaggg}
KLOE coll., A.~Aloisio {\it et al.}, Phys. Lett. B {\bf 591} (2004) 49

\bibitem{etaetap}
KLOE coll., A.~Aloisio {\it et al.}, Phys. Lett. B {\bf 541} (2002) 45

\bibitem{Along} 
A.~Alavi-Harati {\it et al.}, Phys. Rev. Lett. {\bf 88} (2002) 181601

\bibitem{Vud}
A.~Czarnecki, W.~Marciano, A.~Sirlin, [hep-ph/0406324]

\bibitem{KTEV}
T.~Alexopoulos {\it et al.}, [hep-ex/0406003]

\bibitem{LeutRoos}
H.~Leutwyler, M.~Roos, Zeit. f. Physik C {\bf 25} (1984) 91

\end{thebibliography}
\end{document}